\documentstyle[aps,epsfig]{revtex}
\preprint {IMSc-98/01/01}
\begin{document} 
\twocolumn[\hsize\textwidth\columnwidth\hsize\csname @twocolumnfalse\endcsname
\textwidth 16.5cm
\textheight 21.0cm 
\draft
\title{Quantum corrections to the thermodynamic potential of interacting 
Bosons in a trap} 
\author{Subhasis Sinha}

\address
{The Institute of Mathematical Sciences, Madras 600 113, India.}
\date{\today}
\maketitle

\begin{abstract}
We calculate the quantum corrections of the thermodynamic quantities of a 
system of confined Bosons at finite temperature. Systematically quantum 
corrections are written in a series of $\hbar$, which is convergent when 
$kT$ is much larger than the spacing between energy levels of the system. We 
apply this 
method to calculate analytically the thermodynamic potential of a weakly 
interacting Bose-gas confined in 3-d harmonic oscillator potential. For 
large number of particles, quantum corrections become small, and 
contribute to the finite size corrections to scaling.

%\end{abstract}

\end{abstract}
 ]

\narrowtext
%\section{Introduction}

There has been renewed interest in the Bose-Einstein condensation(BEC)
after its experimental demonstration by Anderson et al\cite{anderson}.  It
has now become possible to measure the relevant thermodynamic quantities
of a system of weakly interacting Bosons in a magnetic trap
\cite{ensher,mews}. In these experiments, the temperature dependence of
the condensate fraction and the release energy are measured.
Theoretically, the thermodynamic properties of homogeneous Bose gas was
studied by several authors starting from London\cite{london}. The 
thermodynamic
properties of the inhomogeneous Bose gas was studied by Legget et al
\cite{leggett,oliva}. After the recent experimental 
developments, 
there is a new impetus to understand the thermodynamic properties of
interacting inhomogeneous Bose gas. Within the local density
approximation(LDA), Stringari et al\cite{stringari} have calculated the
thermodynamic quantities numerically and shown scaling properties of 
strongly interacting bosons in a 
trap and since then several calculations have been reported by various
authors\cite{shi,ravndal}. In these papers the thermodynamic properties of
inhomogeneous gas has been calculated within the LDA, neglecting the
quantum corrections which may be important. 

In this paper we develop a systematic semiclassical $\hbar$-expansion
for thermodynamic quantities of an interacting bose gas which takes into 
account the quantum
corrections.  We follow a method analogous to the extended Thomas-Fermi 
method(ETF) which has been successfully applied in the case of finite 
fermion systems, like nuclei, atoms, clusters etc.  It is well known 
that the semiclassical expansion of the smooth part of the quantum 
density of states can be expanded in powers of $\hbar$.   The leading 
term in this expansion is the Thomas-Fermi level density. Systematic 
$\hbar$ corrections to the leading approximation may then be obtained 
using the well known Wigner-Kirkwood(WK) 
expansion\cite{wigner}. In the case of finite 
fermion systems this is what is usually referred to as the extended 
Thomas-Fermi approximation\cite{brack}. 

%\section{Extended Thomas-Fermi approximation for Bosons}
We first develop a systematic $\hbar$ expansion using the W-K 
method\cite{wigner}.
Consider the single particle Hamiltonian $H$ for a system of 
particles, which satisfies the equation,
\begin{equation}
\it{H}\psi_{n} = E_{n}\psi_{n}
\end{equation}
where $E_{n}$ are eigenvalues and $\psi_{n}$ are corresponding wave 
functions. The canonical partition function of the system is defined as 
\begin{equation}
Z_{0}(\beta) = \sum_{n} e^{-\beta E_{n}},
\end{equation}
where $\beta$ is, for our purposes, a parameter with the dimension of 
inverse energy. We first briefly review the method as applied to the finite 
fermion system\cite{brack}.
To extract thermodynamic quantites of the finite fermion system from the 
canonical partition function, a weight function is defined as $w_{F} = 
\frac{\pi\beta T}{\sin(\pi\beta T)}$. By using this weight function one 
can write the grand potential of the system in the following way,
\begin{eqnarray}
kTq(T,z) &=& \mu N -E +T S\nonumber\\ & = & {\sl 
L}_{\mu}^{-1}[\frac{Z_{0}(\beta) 
w_{F}}{\beta^2}]. 
\end{eqnarray}
In integral representation,
\begin{equation}
kTq(T,z) = \frac{1}{2\pi i}\int_{\epsilon -i\infty}^{\epsilon +i\infty}
e^{\beta \mu}\frac{Z_0(\beta) w_{F}}{\beta^2} d\beta
\end{equation}
where the contour is closed from left, enclosing all the negative poles. 
Summing all residues we obtain the known result\cite{pathria},
\begin{equation}
q(T,z) = \sum_{n} (1 + z e^{-E_{n}/kT}),
\end{equation} 
where $z = e^{\beta \mu}$ is the fugacity.
We may extend this method to Bosons also, by taking a suitable weight 
function,
\begin{equation} 
w_{B} = \frac{-\cos(\pi\beta T) \pi\beta T}{\sin(\pi\beta T)}.
\end{equation} 
The grand potential of an ensemble of bosons can be written in terms of 
the following integral,
\begin{equation}
kTq(T,z) = \frac{1}{2\pi i} \int_{\eta -i\infty}^{\eta +i\infty} 
 e^{\beta \mu} \frac{Z_0(\beta) w_{B}}{\beta^2} d\beta.
\end{equation}
The line of integration is choosen in such a way that all the positive 
poles(excluding zero) are to the right side of the line. Since the 
chemical potential
$\mu<E_{n}$ for bosons, we close the contour from right, and summing 
all the residues at the positive poles inside the contour, we obtain the 
well known result\cite{pathria}, 
\begin{equation}
q(z,T) = -\sum_{n} log(1 - ze^{-E_{n}/kT}).
\end{equation}
From the above thermodynamic potential we can calculate all thermodynamic 
quantities.

To do a systematic $\hbar$ expansion of Bosons, we start from high 
temperature limit. Neglecting zero point energy and the occupation number 
of ground state, we can do W-K expansion of canonical 
partition function. For the semiclassical expansion of the partition 
function, it is most convenient to take plane wave basis for calculation 
of the canonical partition function.
\begin{equation}
Z_{0}(\beta) = \frac{1}{h^3}\int d^3p \int d^3r 
e^{-i\vec{p}.\vec{r}/\hbar} 
e^{-\beta H} e^{i\vec{p}.\vec{r}/\hbar}
\end{equation}
To evaluate the partition function, we write,
\begin{eqnarray}
u(\vec{r}, \vec{p}, \beta) & = & e^{-\beta H} 
e^{i\vec{r}.\vec{p}/\hbar}\nonumber\\ & = & 
e^{-\beta H_{cl}} e^{i\vec{r}.\vec{p}/\hbar} w(\vec{r}, \vec{p}, \beta),
\end{eqnarray}
where $H_{cl}$ is the classical hamiltonian $H_{cl} = \frac{p^2}{2M} 
+V(\vec{r})$ and $u$ obeys the Bloch equation,
\begin{equation}
\frac{\partial u}{\partial \beta} + {\sl H}u = 0
\end{equation}
with the boundary condition
\begin{equation}
\lim_{\beta \rightarrow \infty} u = e^{i\vec{p}.\vec{r}/\hbar}. 
\end{equation}
Substituting $u$ in the above equation, we obtain the following equation 
for $w$.
\begin{eqnarray}
\frac{\partial w}{\partial \beta} = -i\hbar 
[\frac{\beta}{M}(\vec{p}.\nabla V)w - \frac{1}{M}(\vec{p}.\nabla 
V)] + \nonumber \\
\frac{\hbar^2}{2M}[\beta^2 (\nabla V)^2 w - \beta(\nabla^2 V)w + 
\nabla^2 w - 2\beta(\nabla V. \nabla w)]
\end{eqnarray}
The above equation can be solved order by order in $\hbar$, by expanding 
$w$ as a power series in $\hbar$.
\begin{equation}
w = 1 + \hbar w_{1} + \hbar^2 w_{2} +...
\end{equation} 
Upto the order $\hbar^2$, canonical partition may then be written 
as\cite{brack}, 
\begin{eqnarray}
Z_{wk}(\beta) & = & \frac{1}{(2\pi\hbar)^3} \int d^3 p 
e^{-\beta\frac{p^2}{2M}} \int d^3 r e^{-\beta V(\vec{r})}\nonumber \\& 
&\times [1 -\frac{\beta^2 \hbar^2}{24 M}\bigtriangledown ^2 V(\vec{r})],
\end{eqnarray} 
where $V(r)$ is the effective single particle potential.
Inserting this expression for canonical partition function in equation 
(7), we obtain the grand potential of the normal state. 
\begin{eqnarray}
q(z,T) & = & -\frac{1}{(2\pi\hbar)^3}\int d^3p \int d^3r 
[log(1-ze^{-\beta(\frac{p^2}{2M} +V(\vec{r}))})\nonumber\\
& & -\frac{\hbar^2}{24M} 
\nabla^{2} V(\vec{r})\frac{\partial^2}{\partial\mu^2} 
log(1-ze^{-\beta(\frac{p^2}{2M} +V(\vec{r}))})]
\end{eqnarray}
After doing the p integration we get the following form of the density of 
the grand potential upto order $\hbar^2$. 
\begin{equation}
F(\vec{r}) = \frac{1}{\lambda^3} [g_{5/2}(\tilde{z}) - 
\frac{\hbar^2}{24M} \nabla^{2} V(\vec{r}) 
\frac{\partial^2}{\partial\mu^2}g_{5/2}(\tilde{z})] 
\end{equation}
where the $g_{n}(x)$ and $\lambda$ are defined as,
\begin{equation}
\lambda = \frac{2\pi\hbar}{\sqrt{2\pi M kT}}~ ;\hspace{.5cm}
g_{n}(x) = \sum_{i=1}^{\infty} \frac{x^i}{i^n}.
\end{equation}
and the effective fugacity is $\tilde{z} = ze^{-\beta V(r)}$.

To check the correctness of the above formalism, we first apply it to a
system of bosons in 3-d isotropic harmonic oscillator(h.o) confinement.  
Following our method, grand potential of the normal state is given by, 
\begin{equation} 
q(z,T) = (\frac{kT}{\hbar\omega})^3 [g_{4}(z) -\frac{1}{8} 
(\beta\hbar\omega)^2
g_{2}(z)] 
\end{equation} 
From this grand potential we can derive the
number of particles and the energy of the normal state. 
\begin{eqnarray}
N_{e}& = &z(\frac{\partial q}{\partial z})_{T}~~~~\nonumber \\& = &
(\frac{kT}{\hbar\omega})^3 [g_{3}(z) -\frac{1}{8}
(\beta\hbar\omega)^2 g_{1}(z)]\\
U& = &-(\frac{\partial q}{\partial\beta})_{z}~~~\nonumber\\ 
& = & kT(\frac{kT}{\hbar\omega})^3 [3 g_{4}(z) 
-\frac{1}{8}(\beta\hbar\omega)^2 g_{2}(z)]
\end{eqnarray}
Above quantities may also be derived by using density of state of the 
system. Exact canonical partition function of the 3-d h.o is 
given by, 
\begin{equation}
Z(\beta) = \frac{1}{[2\sinh(\frac{\beta\hbar\omega}{2})]^3}.
\end{equation}
Density of states can be calculated by taking inverse laplace transform of 
the canonical partition function with respect to $\beta$. Also the same 
results may be obtained by Eular-Maclaurin summation metod\cite{ravndal}. 
From the above calculation we have seen that, 
thermodynamic quantities may be expanded in a series of a dimensionless 
parameter $\beta\hbar\omega$. This series will converge when 
$\beta\hbar\omega \le 1$. However at low temperatures, $\beta\hbar\omega 
\approx 1$, and Wigner-Kirkwood expansion breaks down. At these temperature, 
only few low 
energy states are occupied, so energy levels can not be taken as continuous.

In the systems which show macroscopic occupation of ground state at 
finite temperature, chemical potential of the system tends to zero. As 
$z\rightarrow 1$, the higher order terms in $\hbar$ expansion show infrared 
divergence. To regulate such divergences fugacity of the system can be 
replaced 
by $ze^{-\Delta E/kT}$, where $\Delta E$ is the natural energy gap of finite
size systems\cite{masut}. This is equivalent to introducing a infrared 
momentum cutoff $\approx 1/L$, where $L$ is the system size.

%\section{Application to weakly interacting Bose-gas}
Having checked the formalism, we now apply above method to real 
physical situation that of a non ideal 
Bose gas. In recent experiment Ensher et al\cite{ensher} measured the 
thermodynamic quantities of 40000 $Rb^{87}$ atoms, for which s-wave 
scattering length is $\approx 100a_0$, where $a_0$ is Bohr radius.

Within Hartree-Fock theory of bosons, 
non-condensate part satisfies the following Schr$\ddot{o}$dinger's 
equation\cite{leggett}, 
\begin{equation}
[-\frac{\hbar^2}{2M}\nabla^2 +\frac{1}{2}M\omega^2 r^2 
+2u\rho(r)]\psi_{i}(r) = e_{i}\psi_{i}(r),
\end{equation}
and the condensate part satisfies Gross-Pitaevskii equation,
\begin{eqnarray}
& & [-\frac{\hbar^2}{2M}\nabla^2 +\frac{1}{2}M\omega^2 r^2 
+2u\rho_{nc}(r) +u\rho_{c}(r)]\phi(r)\nonumber\\  & = & \epsilon_{0}\phi(r),
\end{eqnarray}
where $\rho_{nc}(r)$ and $\rho_{c}(r)$ 
are the densities of non-condensate and condensate part, and $\rho = 
\rho_{c} + \rho_{nc}$. The interaction strength is given by,
$u=\frac{4\pi\hbar^2 a}{M}$. Neglecting the kinetic 
energy of the condensate, the lowest energy eigenvalue may be 
approximated by, 
\begin{equation}
\epsilon_{0} \approx \frac{1}{2}\hbar\omega (15N_{0}\frac{a}{l})^{2/5} 
+2u\rho_{nc}(0).
\end{equation}
Further within the same approximation condensate density can be written as,
\begin{equation}
\rho_{c}(r) = \frac{1}{u}(\epsilon_{1} - V(r)),
\end{equation}
where $\epsilon_{1} = \frac{1}{2}\hbar \omega (15 N_{0} \frac{a}{l})^{2/5}$.
The above approximation is valid for large number of condensate atoms 
$N_{0}$ 
and for strong repulsive interaction. This approximation breaks down 
within a small region near the critical temperature, where the number of 
condensate atoms become very small. 
For the non-condensate part we may use the formalism developed earlier. 
Using equation (17), the local thermodynamic potential of the 
non-condensate atoms can be written as,
\begin{equation} 
F(r) = \frac{1}{\lambda^3}[g_{5/2}(\tilde{z}_{eff})
-\frac{\hbar^2}{24M}\nabla^{2}V_{eff} g_{1/2}(\tilde{z}_{eff})],
\end{equation} 
where $\tilde{z}_{eff} = e^{\beta(\mu - V_{eff}(r))}$ and $
V_{eff}(r) = V(r) + 2u\rho(r)$. For perturbative expansion we write the
effective fugacity in the following form. 
\begin{eqnarray} 
\tilde{z}_{eff} & = &
\tilde{z}exp[\beta(\epsilon_1 - V(r) -2u\rho_{c}(r)\nonumber\\& & 
+2u(\rho_{nc}(0) -\rho_{nc}(r)))] 
\end{eqnarray}
where, $\tilde{z} = e^{\beta(\mu -
\epsilon_0)}$, which is always less than 1. Assuming weak coupling, we
expand the functions in terms of $2u(\rho_{nc}(0) - \rho_{nc}(r))$. The 
perturbative expansion in this method turns out to be well behaved\cite{shi}. 
Excluding the non-condensate part, the effective fugacity can be written 
as, 
\begin{eqnarray} 
\tilde{z}_{eff} &=& \tilde{z} e^{\beta(V(r) -
\epsilon_{1})}~~~for~ r<r_0\nonumber\\ & = &\tilde{z}e^{\beta(\epsilon_{1}
- V(r))}~~~for~ r>r_0, 
\end{eqnarray} 
where $r_0$ is the turning point of Thomas-Fermi
density $\rho_{c}$. From the above expression we can see there is no 
expansion parameter for condensate part. We approximate the local 
chemical potential within the condensate by its mean value. To calculate the 
free energy of the system within mean field
theory, we subtract self energy from the total energy of the system. The self
energy contribution is given by, 
\begin{equation} 
E_{s} = u \int \rho_{nc}^{2}(r) d^3 r. 
\end{equation} 
In the following calculation we replace the
coupling by a dimensionless parameter $\eta =
(\sqrt{\frac{2}{\pi}}\frac{a}{l}N^{1/6})^{2/5}$. Upto order $\eta^{5/2}$ and 
$\hbar^2$ the thermodynamic potential is given by, 
\begin{eqnarray}
& & q(z,T)/N = t^3 \tilde{g}(4,\tilde{z},1/2,\alpha \eta \tilde{\beta})
+\nonumber\\ & & \frac{2}{\sqrt{\pi}} t^{3}\sqrt{(\alpha \eta 
\tilde{\beta})}[
g_{7/2}(\tilde{z}) +\frac{2}{3}\alpha \eta \tilde{\beta}~
g_{5/2}(\tilde{z}e^{-2/5\alpha\eta\tilde{\beta}})]~~~~\nonumber\\ & & 
+2\eta^{5/2} t^{7/2}[\zeta(3/2) g_{3}(\tilde{z}) -
\frac{1}{2} F(3/2,3/2,3/2,\tilde{z})]~~~~\nonumber\\ 
& & -\frac{1}{8 N^{2/3}} t[\tilde{g}(2,\tilde{z},1/2,\alpha \eta 
\tilde{\beta}) +\nonumber\\ & & \frac{2}{\sqrt{\pi}}\sqrt{(\alpha \eta 
\tilde{\beta})}
[g_{3/2}(\tilde{z}) - \frac{2}{3}\alpha\eta 
\tilde{\beta} g_{1/2}(\tilde{z}e^{-2/5\alpha\eta\tilde{\beta}})]\nonumber\\ 
& & +2\eta^{5/2} 
t^{1/2}(\zeta(3/2) g_{1}(\tilde{z}) -
F(1/2,-1/2,5/2,\tilde{z}))],
\end{eqnarray}
where the new functions and parameters are defined by,
\begin{eqnarray}
\tilde{g}(s,x,t,y) = \sum_{i=1}^{\infty}\frac{x^i}{i^s}I(t,iy)\\
I(t,y) = \frac{1}{\sqrt{\pi}}\int_{0}^{\infty}dx e^{-x} (x + y)^{(t-1)}\\
F(\alpha,\beta,\gamma,x) = 
\sum_{i=1}^{\infty}\sum_{j=1}^{\infty}\frac{x^{(i+j)}}{i^{\alpha} 
j^{\beta} (i+j)^{\gamma}}\\
\alpha = \frac{1}{2}(15 \sqrt{\frac{\pi}{2}}N_0/N)^{2/5}\\
t= 1/\tilde{\beta} = kT/(\hbar \omega N^{1/3})
\end{eqnarray}
For small value of $\alpha \eta$, the function 
$\tilde{g}(s,\tilde{z},1/2,\alpha\eta\tilde{\beta})$ can be written as 
$g_{s}(\tilde{z})$.
Condensate fraction can be calculated from the thermodynamic potential,
\begin{equation}
N_{0}/N = 1 - z\frac{\partial(q(z,T)/N)}{\partial z}
\end{equation}
For $T<T_{c}$, $\tilde{z}\approx 1$. To regulate some divergent terms 
appearing in the $\hbar^2$ corrections, we set $\tilde{z} \approx 
e^{-\beta\hbar\omega}$, which is the natural energy gap from ground state.
From the above expression we can see that the condensate fraction and other 
thermodynamic quantities are not fully scale invarient with respect to 
the dimensionless scaling parameters $\eta$ and $t$. But the extra terms 
coming from  the quantum corrections contain a factor $1/N^{2/3}$. These 
terms depend on the number of particles in the system, and give finite size 
corrections to the scaling form of the thermodynamic quantities. In the 
large $N$ limit and also in the high temperature phase, quantum corrections 
are negligible. But for small quantum confined systems they give 
finite corrections to the free energy.

\begin{figure}[htb]
\begin{center}
\makebox{\epsfig{file=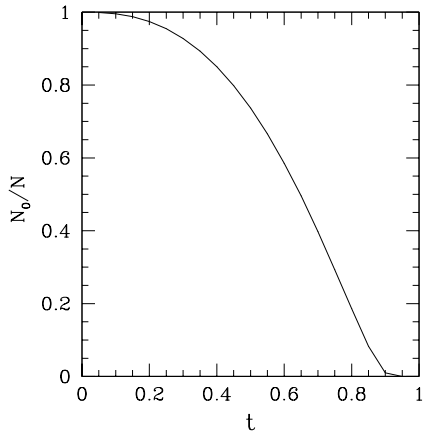,width=6.0cm,height=6.0cm}}
\caption{Condensate Fraction for 10,000 atoms, $\eta$=.18.}
%\label{fig:confrac}
\end{center}
\end{figure}

\begin{figure}
\begin{center}
\mbox{\epsfig{file=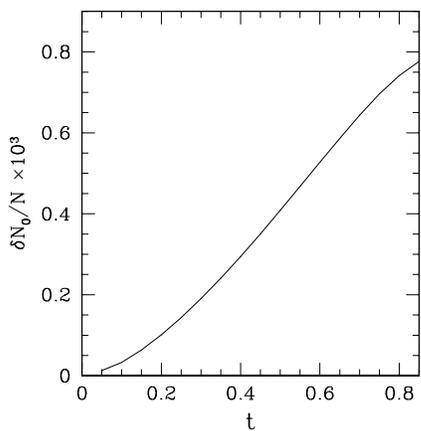,width=6.0cm,height=6.0cm}}
\caption{Finite size correction to condensate fraction for 10,000 
atoms, where $t$ is reduced temperature.} 
%\lebel{fig:qcorr}
\end{center}
\end{figure}

For real physical systems the typical value of the parameter $\eta$ which 
determines the two-body interaction strength varies from $0.1 -0.25$. In 
Fig.1 we have shown the variation of condensate fraction with reduced 
temperature $t$, by choosing $N = 10,000$ atoms and the dimensionless 
parameter $\eta = 0.18$. In Fig.2 we have shown the finite size 
correction to the condensate fraction for same values of the parameters. 
From the graph we can see that the first order quantum correction is two 
order smaller than the leading order contribution.

In conclusion, we have developed an extended Thomas-Fermi method at finite 
temperature for a system of Bosons. Systematic quantum corrections may be 
written in a series of a dimensionless parameter $\beta \hbar\omega$, 
which converges for the temperatures larger than the average spacing 
between energy levels. We have applied our method to a system of 
weakly interacting Bose gas confined in isotropic harmonic oscillator 
potential. Thermodynamic potential and condensate fraction are calculated 
analytically upto $\eta^{5/2}$ and $\hbar^2$ order. Leading terms in 
$\hbar$ are the functions of scaling parameters $t$ and $\eta$ only, but 
leading order quantum corrections give $N$ dependent corrections to the 
scaling form. In present situation quantum corrections are negligibily 
small. But quantum corrections will be important for small number of 
particles and for low dimensional systems. Also its magnitude depends on 
the nature of confinement potential. This method may be applied to 
investigate the properties of a system of confined charged bosons in weak 
magnetic field.

I would like to thank M.V.N. Murthy and R.K. Bhaduri for helpful discussions.


\bigskip 

\begin{references}
 
\bibitem{anderson} 
M. H.Anderson, J. R. Ensher, M. R. Matthews, C. E. Wieman, and E. A.  Cornell,
Science {\bf 269}, 198 (1995). 

\bibitem{ensher}
J. R. Ensher, D. S. Jin, M.R. Matthews, C.E. Weiman and E. A. Cornell, 
Phys. Rev. Lett. {\bf 77}, 416(1996).

\bibitem{mews}
M. O. Mews, M.R. Andrews, N.J. van Druten, D.M. Kurn, D.S. Durfee, and W. 
Ketterle, Phys. Rev. Lett. {\bf 77}, 416(1996).

\bibitem{london}
F. London, Nature {\bf 141}, 643(1938) ; L. Landau, J.phys. USSR {\bf 5}, 
71(1941).

\bibitem{leggett}
V. V. Goldman, I. F. Silvera and A. J. Leggett, Phys. Rev.B {\bf 24},
2870(1981).

\bibitem{oliva}
J. Oliva, Phys. Rev. B {\bf 39}, 4197(1989).

\bibitem{stringari}
S.Giorgini, L. P. Pitaevskii, and S. Stringari, Phys. Rev. Lett. {\bf
78}, 3987(1996).

\bibitem{shi}
H. Shi and W. M. Zheng, Phys. Rev. A {\bf 56}, 2984(1997).

\bibitem{ravndal}
H. Haugerud, T. Haugset and F. Ravndal, Phys. Lett. A {\bf 225}, 18(1997).

\bibitem{wigner}
E. Wigner, Phys. Rev. {\bf 40}, 749(1932) ; J. G. Kirkwood, Phys. Rev.
{\bf 44}, 31(1933).

\bibitem{brack}
Matthias Brack and Rajat. K. Bhaduri, Semiclassical Physics,
(Addison-Wesley publishing company, 1997).

\bibitem{pathria}
R. K. Pathria, Statistical Mechanics, (Pergamon Press, New York, 1972).

\bibitem{masut}
R. Masut and W. J. Mullin, Am. J. Phys. {\bf 47}, 493(1979).

\end{references}
\end{document}